\pdfoutput=1

\documentclass{svproc}
\usepackage{url}
\usepackage{soul} 

\usepackage{enumitem}
\usepackage{changepage}

\newcommand{\iv}{\mbox{IV}}
\newcommand{\codestyle}{\texttt}
\newcommand{\ilspycmd}{\codestyle{ilspycmd}}

\newtheorem{defn}{Definition}

\begin{document}
\mainmatter             
\title{Source Code Anti-Plagiarism: a C\# Implementation using the Routing Approach}
\titlerunning{Source Code Anti-Plagiarism}  
%
\author{Fabrizio d'Amore \and Lorenzo Zarfati}
\authorrunning{F. d'Amore and L. Zarfati} 
%
\tocauthor{}
\institute{Sapienza University of Rome, Italy,\\
\email{damore@diag.uniroma1.it \\ zarfati.1598154@studenti.uniroma1.it} 
}

\maketitle              

\begin{abstract}
Despite the approaches proposed so far, software plagiarism is still a problem which hasn't been solved entirely yet. The approach introduced throughout this paper is about a source code anti-plagiarism technique which aims at rendering the source code incomprehensible to a possible plagiarist and at the same time preventing source code modifications. The proposal is based on the concept of \emph{Router} and makes use of both symmetric encryption and cryptographic hashing functions to provide such guarantees.

\end{abstract}
\section{Introduction}
In general plagiarism can be defined as the act of taking someone else's work or idea, and pass it as your own without giving the proper credits to the original authors. 
It can affect anything that has a value including academic research papers, intellectual property and software products, the last of which is causing huge money losses to companies worldwide. Naturally it is not limited to copy-paste activities as they are the technique which is the most trivial to be detected. There exist many plagiarism techniques that might be used depending on the target, for instance mosaic and self-plagiarism are two of the most common in research papers.

However when it comes to software plagiarism, both malicious and countermeasure activities change, and the challenges become even harder when compared with other  domains where it must be considered. Plagiarism was not the main concern when the first software products started to be released, it is a problem which was considered only afterwards, while many products inevitably were affected already.

Our approach relies on the introduction of a software module, which we call ``Router'', that through some suitable cryptographic operations, can make it too difficult to copy or modify relevant parts of the source code. In our model we consider it meaningful the effort spent in the plagiarism: we do not need a provably secure approach but a heuristic that makes plagiarism costly as the process of rewriting the source code.

\subsection{Previous work}
During the past decades several techniques have been proposed to fight software plagiarism even though nowadays the problem is not entirely solved yet. The majority of approaches proposed so far mainly aim at detecting plagiarism once it has occurred, thus when it's too late, rather than preventing its occurrence. The detection techniques that have been the subject of several studies are code clone detectors, watermarking and birthmarking schemes. Code clone detectors are software used to detect code fragments which are considered the be equal according to a well-defined similarity definition. Initially introduced as an optimization technique for shortening source files sizes  and incentivize abstraction, they have also been used to detect plagiarism in source code \cite{codeclone} and nowadays they are mostly used for binary similarities detection, dealing with assembly instructions rather than high-level programming languages each with its own syntax. The trend is increasingly relying on ML-based approaches which turned out to provide astonishing improvements, especially regarding performance \cite{siamese}.

Watermarking schemes were initially proposed to protect copyright of digital documents such as images and only afterwards they were employed to detect software plagiarism. Watermarks can be informally defined as unique identifiers computed over specific software characteristics, which are inserted within software itself and retrieved subsequently to detect partial or entire copy of two software instances. Some of the most effective approaches are those based on dynamic execution tracing \cite{dynamicwat} though most recently proposed techniques rely on concepts such as the Collatz conjecture and the Shamir secret scharing scheme \cite{collatz,shamir}, which will serve as a starting point for future studies. 

On the other hand, birthmarking schemes do not modify the software code but rather they are based on one or more intrinsic characteristics of the software itself. This kind of schemes is typically used to identify a program behavior even though it was modified successfully or included within another bigger software fragment. Since they do not modify the software, they can only be used to state whether one program is likely to be a copy of another. They are stealthier than watermarking schemes even though the former cannot be used to prove the owner of a software (or digital document) effectively has the right to own it, unless this is achieved by combining the two kind of schemes.

Watermarking and birthmarking schemes are usually classified as either static or dynamic, depending on whether their computation is based on static data only, which are accessible without executing the software, or if it is based on dynamic software state which can be reached only at runtime.
Since nowadays no foolproof solution exist for either scheme, several researches in the literature recommends to combine both the schemes and especially static and dynamic techniques. 
Note that all the mentioned techniques allow to detect plagiarism rather than attempting to prevent its occurrence altogether. The only technique not already mentioned which can really help in preventing source code plagiarism is code obfuscation. By obfuscation we mean any semantic-preserving transformation which renders the code unreadable to a potential plagiarist without altering the software functionalities. Unless the obfuscation algorithm is reversed, software is made intelligible therefore preventing a potential plagiarist from using it for its own purpose, either entirely or just partially \cite{obfuscation}. 

The proposed technique partially relies on obfuscation but together with symmetric encryption and cryptographic hash functions aims at preventing plagiarism attempts rather than just detecting them once they were carried out successfully.

The rest of this paper is organized as follows: the following section describes the approach underlying our prototype; Section~\ref{se:implementation} and \ref{se:testSection} provide implementation details and tests outcomes respectively; Section~\ref{se:secCons} provides security considerations about both proposed approach and the C\# implementation, whereas the last one is about conclusions and future works. 

\section{A new approach}\label{sec:newapproach}

The idea over which the proposed approach is based on an idea that has been investigated for several years, without however publishing the preliminary results. Terrinoni was the first master student to write a thesis \cite{terrinoni} on it; later the idea was refined. It is inspired by the behavior of a well-known network device: the router, which is responsible of correctly routing packets depending on their destination address. The behavior is quite similar since the code implementing the approach consists in forwarding function references containing encrypted parameters toward the intended destination function and then return the output of such function (if any) to the original caller. 
From now on we will refer to the code responsible of performing all the features described in this paper with the term \emph{Router} to make a clear distinction between the common networking device and the code implementing the approach.

The approach consists in rewriting function references (calls) to be protected with correspondent Router references, according to well-defined format which will be detailed in Section~\ref{se:implementation}. Router calls input parameters will be encrypted by the Router itself during its initialization step and once this step terminates, and the software is re-compiled considering modified source files, encrypted parameters are decrypted only at runtime by the Router code which will then invoke the original function if and only if some particular conditions are met. It's very clear that this approach modify the software source code to be protected and it does so in only two ways: function calls replacement, which can be performed during the development itself or subsequently; the \emph{Router.Init()} invocation must be inserted as the first statement of the software to be protected, so that a proper Router initialization step can be performed. The approach mainly relies on static features but it also depends on a specific runtime program state, therefore it may be classified as a blended technique. Before describing the approach further the definition of file dependency and closed hash values have to be clear.

\begin{defn}[file dependency]
A source file $f_i$ depends on another source file $f_j$  ($f_i \to f_j$) if and only if (functions in) $f_i$ reference (call) a function defined within source file $f_j$ and both files are part of the same software.
\end{defn}

The definition does not consider references to files belonging to different software nor self-dependencies, the latter are not considered because they are also some of the most common. These considerations are made as a trade-off between security and performance: replacing all the function references might lead to severe performance degradation even though there might be scenarios where this becomes viable.

\begin{defn}[closed hash]\label{def:closed_hash}
Given a DAG $G = (V, A)$ where $V$ is the set of source files that contain functions called by different source files, and each arc $(f_i, f_j) \in A$ is a file dependency: some function of $f_i$ calls some function of $f_j$. The closed hash value $h_i$ of the $i$-th file having $m$ dependencies $f_{i_1}, \ldots, f_{i_m}$, is defined as:

$$h_i = h_i' \oplus h_{i_1} \oplus h_{i_2} \oplus \cdots \oplus h_{i_m}$$
\noindent where:
\begin{itemize}
  \item $h_i'$ is the traditional hash value computed just on file $f_i$ ($h_i' = h(f_i)$, being $h$ the chosen cryptographic hashing function;
  \item $h_{i_1}, h_{i_2}, \ldots, h_{i_m}$ are the closed hashes of files $f_{i_1}, \ldots, f_{i_m}$ which file $f_i$ depends upon.
\end{itemize}
If the file $f_k$ has no dependency, then the closed hash is its standard (traditional) hash ($h_k = h_{k}' = h(f_k)$).
\end{defn}
We point out that while talking of dependencies we consider only the original source code (not including the Router). In the first manuscript \cite{terrinoni} the closed hashes were called ``solved hashes'', and thus it has been done in subsequent manuscripts \cite{cardelli,cavallaro}.

Closed hash value computation for a given file might lead to a deadlock situation if there exist some circular dependency among source files, or equivalently, if one or more cycles exist in the directed graph obtained as described above (in Def.~\ref{def:closed_hash}). Therefore when computing closed hash values the Router has to consider source files, thus graph vertices, in a well-defined order to prevent any deadlock scenario. It can be obtained by a topological sort and it is well-known that a topological sort exists if and only if the graph is acyclic. For this reason, first the Router runs a deterministic algorithm to break every cycle (if any) arbitrarily. Then, the topological sort is done (deterministically, again) and then closed hash values can be computed according its order (backward). These mentioned values are computed by the Router during its initialization step, together with random \iv{}s (Initialization Vectors), one for each file that has information to be encrypted. Such keys are obtained xor-ing the xor of all the closed hashes of dependent files with a nonce (the \iv) for preventing any known-ciphertext attack. 

Of course the Router is a function written in another file. Let us briefly summarize the process:
\begin{enumerate}
  \setcounter{enumi}{-1} 
  \item The original source code has been produced but not yet secured by our approach;
  \item all calls to functions in different files are replaced by a call to the Router, with parameters the name of the originally called function, and a list of comma separated original parameters;
  \item the Router is called, for pre-processing the software and doing some verifications; in particular it symmetrically encrypts by using a proper  key (see details later) names and parameters that describe the original calls (and changes itself!);
  \item the code of the Router is obfuscated, in order to hide the logic of the original program;
  \item the software is now protected and can be released and deployed.
\end{enumerate}

The symmetric encryption key is computed only during the initialization step and a second time at runtime, when the software needs it to decrypt data previously encrypted. Since the key is never stored within the software it prevents easy key extractions from the software source code leaving a potential plagiarist with no option but performing a dynamic analysis with the aim to infer the Router behavior and remove all the modifications performed before the software was released. 	

The symmetric encryption key computation process is summarized below:

\begin{equation}\label{eq:key}
k_j = h_{j_1} \oplus h_{j_2} \oplus ... \oplus h_{j_n} \oplus \iv_{j}
\end{equation}
where:
\begin{itemize}
  \item $k_j$ is the key used to encrypt function call information of functions belonging to $f_j$ coming from other files (but now from the Router);
  \item $h_{j_1}, h_{j_2}, \ldots, h_{j_n}$ represent all the closed hashes of files which depend on $f_j$ (destinations of the outgoing arcs);
     \item $\iv_{j}$ is a nonce generated during Router initialization, associated with the $j$-th file ((part of) whose functions should be called by external files, sources of the incoming arcs).
\end{itemize}

Thus the key computation process depends on closed hash values and since such values are always the same as soon as the same input DAG is provided, then it follows the key uniqueness theorem \cite{terrinoni}. 
However, note that both the encryption key and closed hashes must be computed over decompiled source files rather than the original ones that, for instance, contain user comments and blank lines.
This is strictly required because once the software is compiled and released, the Router code will have no way to access the original source files and therefore any hash value on different files will never match the one computed during the Router initialization step. 

Once closed hash values and \iv{}s are computed they are printed to standard output, together with file dependency info as Base-64 encoded strings which have to be copied within the Router very first lines of code, replacing homonymous declared yet not defined variables. This step can be easily automated and even though it is planned for future releases the implemented technique currently leaves this task to developers.

Now that the initialization step has been described in more detail it is time to describe how the Router detects plagiarism at runtime. Every reference's first parameter to Router code is encrypted and contains all the data required to invoke a specific function, such as its name and parameters as well as the name of the enclosing class. Since such parameter was previously encrypted with a key depending on files (according to Equation~(\ref{eq:key})), it follows that as soon as the decryption succeeds we know such files were not modified (we currently use \codestyle{aes-256-cbc}, but we plan to switch to the more secure authenticated encryption\footnote{A method that guarantees confidentiality and authentication. Beginners can see, e.g., \cite{10.1145/3131276,DBLP:journals/jzusc/ZhangLYZGR18}.}). On the other hand before invoking the original function, the Router ensures that runtime computed closed hash value of the calling file corresponds to the same value computed during the initialization step. If these values are equal and a secure cryptographic hash function is used, such as one of the SHA-2 or SHA-3 family, then we also know that any potentially involved file was not altered in any way. Thus if either the decryption of the string parameter fails or if any runtime computed closed hash does not match the previously computed one, then the Router terminates the running software and displays a message to the user indicating a plagiarism attempt was detected. 

It is important to highlight that this is not the only response that might be provided by the Router. In general, any kind of task may be performed such as the deletion of the involved software from the target computer. Alternatively, the Router might also inform the original authors or the law enforcement themselves once or incrementally, every time one of the mentioned conditions is not met at runtime. To be precise the aim of the proposed approach is to perform something that inevitably lead the plagiarism attempt to be stopped. However, these additional tasks were not considered in the C\# implementation as they depend on both the software and the environment involved. Thus this additional step was left as a feature to be implemented in future releases. The next section briefly describes implementation details and the steps required to protect a executable or DLL file obtained by compiling one or more C\# source files.

\section{C\# Implementation}\label{se:implementation}
The Router functionalities have been implemented through multiple C\# classes which can be easily imported since they are part of a common C\# VS (Visual Studio) project. We chose to use C\# as it is usually the preferred choice when implementing executables and DLL files for Microsoft Windows, which is still the most used operating system. 

\subsection{Software Dependencies}
The Router code uses both reflection and cryptographic API to retrieve/invoke a specific function and to perform encryption and closed hashes related tasks, respectively. Apart these two APIs the software also has two additional dependencies:

\begin{itemize}
  \item \codestyle{Roslyn}. C\# Code Analysis API.
  \item \ilspycmd.\footnote{\url{https://www.nuget.org/packages/ilspycmd/}} A command-line decompiler using the \codestyle{ILSpy}\footnote{\url{https://github.com/icsharpcode/ILSpy/releases/}} decompilation engine.
\end{itemize}

\noindent The set of C\# code analysis API is used while inspecting source files to quickly retrieve references to what we name ``Router \codestyle{forwardCall}.'' More precisely, such API is used to retrieve effective references, avoiding error-prone \codestyle{grep} tasks, and their encrypted first parameter which contains all the data required to invoke a specific function. 

The second dependency is the accurate .NET decompiler used both at runtime and during the initialization to decompile the running software. \ilspycmd\ is one of the most popular public .NET decompiler even though it requires the program to access a command shell and therefore it restricts the usage to CLI-enabled environments such as desktop and server computers. To the best of our knowledge, \ilspycmd\ is the most used .NET decompiler which can be used programmatically whereas other known decompilers are meant to be used interactively through a GUI.

\subsection{Anti-Plagiarism Steps}

Once the Router project is referenced by the VS project(s) of the application source code to be protected, then we follow the steps mentioned below. 

\begin{enumerate}
  \item\label{en:replace} Replace original function calls (only those calling to other files) with correspondent Router \codestyle{forwardCall}.
  \item\label{en:init} Run the Router initialization step.
  \item\label{en:ccompile} Copy output values within the Router code and compile the software with the modified source files.
  \item\label{en:obfuscate} Obfuscate the Router code.
\end{enumerate}

\noindent The calls replacement of the first step must be performed according to a predefined format which essentially consists in invoking a Router \codestyle{forwardCall} instead, with all the parameters provided as a hyphen-separated values string. The mentioned format is shown below whereas an example with dummy objects is shown right after it.

\begin{center}
\codestyle{[dstClssFullName]-[methodName]-[parametersTypes]-[parametersValues]}
\end{center}

\begin{verbatim}
Class1.f1();                 (defined in Class1)
obj.f2();                    (obj is an instance of Class2)
obj.f3(7, "aString", varX);  (obj is an instance of Class3)
\end{verbatim}

\noindent The expected function call replacement is shown below:
{\footnotesize \begin{verbatim}
Router.ForwardCall("Class1-f1-null-null");
Router.ForwardCall("Class2-f2-null-null", obj);
Router.ForwardCall("Class3-f3-System.Int32,System.String-7,aString", obj, X);
\end{verbatim}
}

\noindent Steps~\ref{en:init}.\ and~\ref{en:ccompile}.\ require little or no effort and are fast to be performed whereas the last one might require additional time. The Router code must be obfuscated because otherwise, anyone having access to the executable would decompile it, understands the internal logics and finally renders the protection scheme worthless by decrypting the ciphertext. It is recommended to apply obfuscation to both static and dynamic variables to obtain stronger guarantees (e.g. using SmartAssembly\footnote{https://www.red-gate.com/products/dotnet-development/smartassembly/}). 

All the mentioned steps must be performed sequentially since each of them requires the previous one to be finished and therefore optimization choices were quite limited and were considered for each step individually. For instance, the Router implements a simple methods caching mechanism to prevent retrieving the same function object repeatedly in a short amount of time. The reader in encouraged to refer \cite{zarfati} for additional details.

\section{Tests Outcomes} \label{se:testSection}

The implementation performance impact was first compared with a previously proposed Java prototype \cite{cardelli} and subsequently evaluated on a real world software. First comparison consists in invoking the same function one billion times and the outcomes of such tests are summarized within the next table. 

\begin{center}
\begin{tabular}{||c|c||} 
 \hline
 Technique & avgTime \\ [0.5ex] 
 \hline\hline
 Direct Access & 19s  \\ 
 \hline
 Reflection+Caching & 155s  \\
 \hline
 Reflection & 238s  \\
 \hline
\end{tabular}
\end{center}

\noindent where we call ``Direct Access'' the bypassing of the Router (original unchanged calling style). As expected, the direct access test scenario is the one that required less time to complete whereas the reflection tests spent more time on average with the evident benefit of caching function objects in memory. When compared with the previous Java prototype, the latter apparently performs better according to reported tests outcomes \cite{cardelli}. Although the outcomes are quite similar, the Java LambdaMetafactory\footnote{See, for example, \url{https://docs.oracle.com/javase/8/docs/api/java/lang/invoke/LambdaMetafactory.html}} approach provides astonishing performance at the expense of usability, with a delay closer to direct access than to the one where reflection API is used. Despite this performance difference, there is no mentioned public software implemented using such technique while we provide a usable C\# implementation relying on reflection.\footnote{https://github.com/msc-antiplag}

\noindent Our implementation was also tested on KeePass, one of the most used open-source password manager implemented in C\#. KeePass is composed by more than 400 C\# classes and only those belonging to the utility package was considered during tests, resulting in 200 replaced function references. The time required by the Router initialization step has been 12s on average and, without considering the decompilation delay, it would be shortened to 7s on average only. However, note that also this delay depends on the size of software to be protected. Since the initialization delay is never experienced by the end user and such delay would just be added to the overall time required for releasing a software product, it has been considered quite short. On the other hand, at most 1s delay was experienced by the end user during tests that lasted up to 2 minutes and during which the main application functionalities were used nonstop. Despite tests outcomes were considered quite positive an extensive evaluation is required, especially before using the Router code for software subroutines with tight time constraints. However it is good to remember that experiencing some time delay is unavoidable when compared to direct function references since there are many other tasks to be performed before the intended function is invoked. 

One important downside which has to be mentioned is intrinsic in the approach over which the implementation is based: the runtime decompilation delay, which would be experienced by the user for sure. However \cite{zarfati} this can be somewhat mitigated by displaying a progress bar before running the software, indicating that anti-plagiarism related activities are running. Once again is important to point out that this delay will depend on both the software size (in terms of file number and their sizes) and the time efficiency of the decompiler being used.

\section{Security Considerations}\label{se:secCons}

This section is subdivided in multiple subsections providing security considerations regarding both the proposed approach and its C\# implementation. In particular, they will be discussed the possibilities left to a possible plagiarist by considering that he/she already has access to the software and therefore to its corresponding decompiled source files.

\subsection{Comparison with Java implementation}

From an attacker perspective, once the source code is obtained all that remains to do is to decrypt data and retrieve original function references as if the Router were never been used. 
When compared to previous Java proposal \cite{cardelli}, the C\# implementation provides stronger guarantees in terms of security as it does not leak any useful information to a potential plagiarist. On the other hand, previous Java proposals leak the class name of the class where the function which is going to be invoked is defined, thus rendering the task of an eventual plagiarist far easier. Every reference to the Router \codestyle{forwardCall} function will appear as depicted below, with a longer (non-truncated) Base-64 encoded string as first parameter.

{\footnotesize

\begin{adjustwidth}{-20pt}{-20pt}
\codestyle{Router.forwardCall("wAp48JGP1bYqVzCYiwuNwSIKVA==", "KeePass.Util.BinaryDataUtil");} 
\end{adjustwidth}
}

\smallskip\noindent First encoded parameter corresponds to the encrypted binary string containing all the information required to invoke the original function. Once decoded it provides no useful information to anyone having no access to the decryption key. The second plaintext parameter leaks no information since it's just used to shorten the overall delay of \codestyle{forwardCall} and it might be obtained anyway by just decompiling the running software.

\subsection{No trial-error guessing}

In order to guess the original function references, replaced and encrypted according to proposed approach, a potential adversary might either attempt to break the encryption scheme or guess all the original function references at once. However, decrypting something previously encrypted with AES and 256 bits keys is known to be unfeasible by spending any reasonable amount of time. Thus, what remains him to do is to guess all the functions original references and despite it has been made harder than before it is not impossible to be accomplished. 

The difficulty in undoing all the Router performed tasks lies in the latter hindering any possible trial-error guessing approach. The implementation as it is just terminates the running software and displays a message to the end-user as soon as a plagiarism attempt is detected. However, as mentioned in Section~\ref{sec:newapproach}, other kinds of tasks could be performed, especially those that would inevitably lead a potential plagiarist to stop its attempt, such as the software deletion and incremental alerts sent to proper people.

\subsection{Strongly Typed Programming Languages}

When a plagiarist will attempt to guess original references he will be helped by the nature of the used programming language. Both C\# and Java are considered strongly-typed programming language and informally, as such, each variable has a specific type and anyone with source code access might determine the type of a chosen variable. Furthermore the function output type is known too and therefore, for instance, if the sole function returning a boolean is replaced then it would be trivial to retrieve its original unencrypted reference by looking at the output type of the corresponding encrypted reference to Router \codestyle{forwardCall}. Since the usage of a strongly typed programming language somewhat weaken the C\# implementation it is highly recommended to replace (i.e. protect) only those references to functions for which exist at least one other function definition returning the same output type. Naturally the more references are replaced the more difficult it will be for an adversary to plagiarize the protected software.

\subsection{Dynamic Analysis}

A potential plagiarist cannot understand much about the protected function references and he might just have some intuition about the original ones, together with all its input parameters, by looking at the context in which such reference is found. For instance, it would be trivial to guess the original reference to function returning a string value if the surrounding code performs string manipulation and there exist only two function candidates: one performing string manipulation as well and the other performing network-related activities. Thus it will be a responsibility of the developer, or the cybersecurity analyst, to choose what references should be and what should not be replaced. If this is done properly, then there would not be much information obtainable through a static analysis. However, dynamic analysis can provide additional information. Depending on the obfuscation algorithms used, it might be possible to understand which files are accessed and then infer the technique used to compute both the encryption key and the runtime hash checking mechanism, even though this requires far more skills than a simpler static code analysis and is not guaranteed to succeed. 

\section{Conclusions}
\label{sect:conclusions}
We presented a new approach to prevent the source code plagiarism. Our prototypes showed that the approach is viable, at least while programming in Java and in C\# (for Java see \cite{cardelli,cavallaro}). No other similar approach is known to the best of our knowledge.

We understand that our solution is not unassailable (some advanced schemes of attack are shown in \cite{cardelli}), however the main purpose is to render plagiarism so long, boring and complicated to make it competitive to re-write the software. The Router needs to be obfuscated, in order to prevent its analysis and understanding.

Our method could lead to a new scheme/framework in software production, making the addition of the anti-plagiarism technique integrated in the life cycle of the software product to be protected.

For the future, we are looking at the Javascript language, and this will be the object of our next work. We aim at providing a similar technique to Javascript code, since this language is very used in the construction of web sites/applications. Thus, our approach could prevent the copying of code from some source (a web site) and its reuse in other sites, a practice that is very common on the modern web.

\medskip\noindent
The authors acknowledge the contributions to this research by former master students Terrinoni \cite{terrinoni}, Cardelli \cite{cardelli} and Cavallaro \cite{cavallaro}, who got their master degree in 2018 or 2019, allowed to develop a more mature approach and tested the method for the Java language.

\bibliographystyle{plain}

\bibliography{biblio}

\end{document}